\documentclass{PoSi}
\usepackage{amsmath}

\def\eq#1{\begin{equation}#1\end{equation}}
\def\del{\partial}
\def\fref#1{Fig.\,\ref{#1}}

\graphicspath{{figs/}}

\title{Search for dark matter in the hidden-photon sector with a large
spherical mirror}

\ShortTitle{Search for dark matter with FUNK mirror}

\author{
\speaker{Darko~Veberi\v{c}}$^{\,a}$,
Kai~Daumiller$^{\,a}$,
Babette~D\"obrich$^{\,b}$,
Ralph~Engel$^{\,a}$,
Joerg~Jaeckel$^{\,c}$
Marek~Kowalski$^{\,d\,e}$,
Axel~Lindner$^{\,d}$,
Hermann-Josef~Mathes$^{\,a}$,
Javier~Redondo$^{\,f}$,
Markus~Roth$^{\,a}$,
Christoph~M.~Sch\"afer$^{\,a}$,
Ralf~Ulrich$^{\,a}$
[The~FUNK~Experiment]
\vspace{1mm}
\\
\llap{$^a$} Institute for Nuclear Physics, Karlsruhe Institute of Technology
(KIT), Germany\\
\llap{$^b$} Physics Department, CERN, Geneva, Switzerland\\
\llap{$^c$} Institute for Theoretical Physics, Heidelberg University, Germany\\
\llap{$^d$} Deutsches Elektronen Synchrotron DESY, Hamburg, Germany\\
\llap{$^e$} Department of Physics, Humboldt University, Berlin, Germany\\
\llap{$^f$} Department of Theoretical Physics, University of Zaragoza, Spain\\
\llap{$^*$} E-mail: \href{mailto:darko.veberic@kit.edu}{\rm darko.veberic@kit.edu}
}

\abstract{
If dark matter consists of hidden-sector photons which kinetically mix with
regular photons, a tiny oscillating electric-field component is present
wherever we have dark matter. In the surface of conducting materials this
induces a small probability to emit single photons almost perpendicular to the
surface, with the corresponding photon frequency matching the mass of the
hidden photons. We report on a construction of an experimental setup with a
large ${\sim}14$\,m$^2$ spherical metallic mirror that will allow for searches
of hidden-photon dark matter in the eV and sub-eV range by application of
different electromagnetic radiation detectors. We discuss sensitivity and
accessible regions in the dark matter parameter space.
}

\FullConference{
The 34th International Cosmic Ray Conference\\
30 July -- 6 August, 2015\\
The Hague, The Netherlands
}

\begin{document}

\section{Introduction}

There is a number of convincing astrophysical and cosmological evidences that a
large fraction of the energy density in the universe must be composed of
invisible non-baryonic matter or dark matter (DM)~\cite{gelmini}. The most
explored options for explaining DM are extensions of the Standard Model (SM)
predicting axions and weakly-interacting massive particles. In recent years
attention has been turned also to alternatives or weakly-interacting slim
particles (WISP), as e.g.\ axion-like particles or hidden photons (HP)~\cite{Nelson:2011sf}. WISP
could be nonthermally produced in the early universe and survive as cold DM
until today (see e.g.~\cite{Jaeckel:2010ni,Ringwald:2012cu,Jaeckel:2013ija,baker} for reviews).

\begin{figure}[t]
\centering
\includegraphics[width=0.43\textwidth]{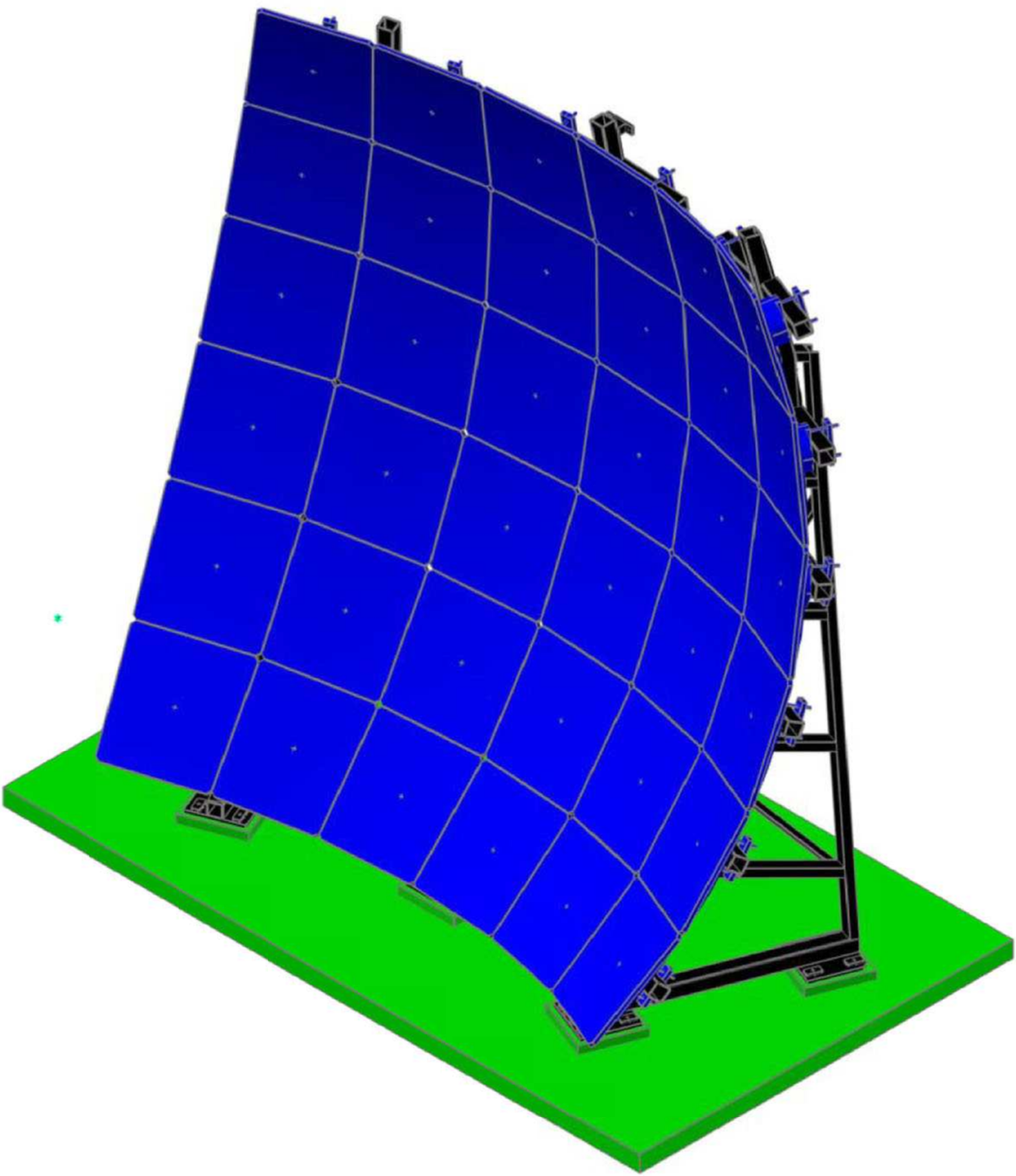}\hfil
\includegraphics[width=0.48\textwidth]{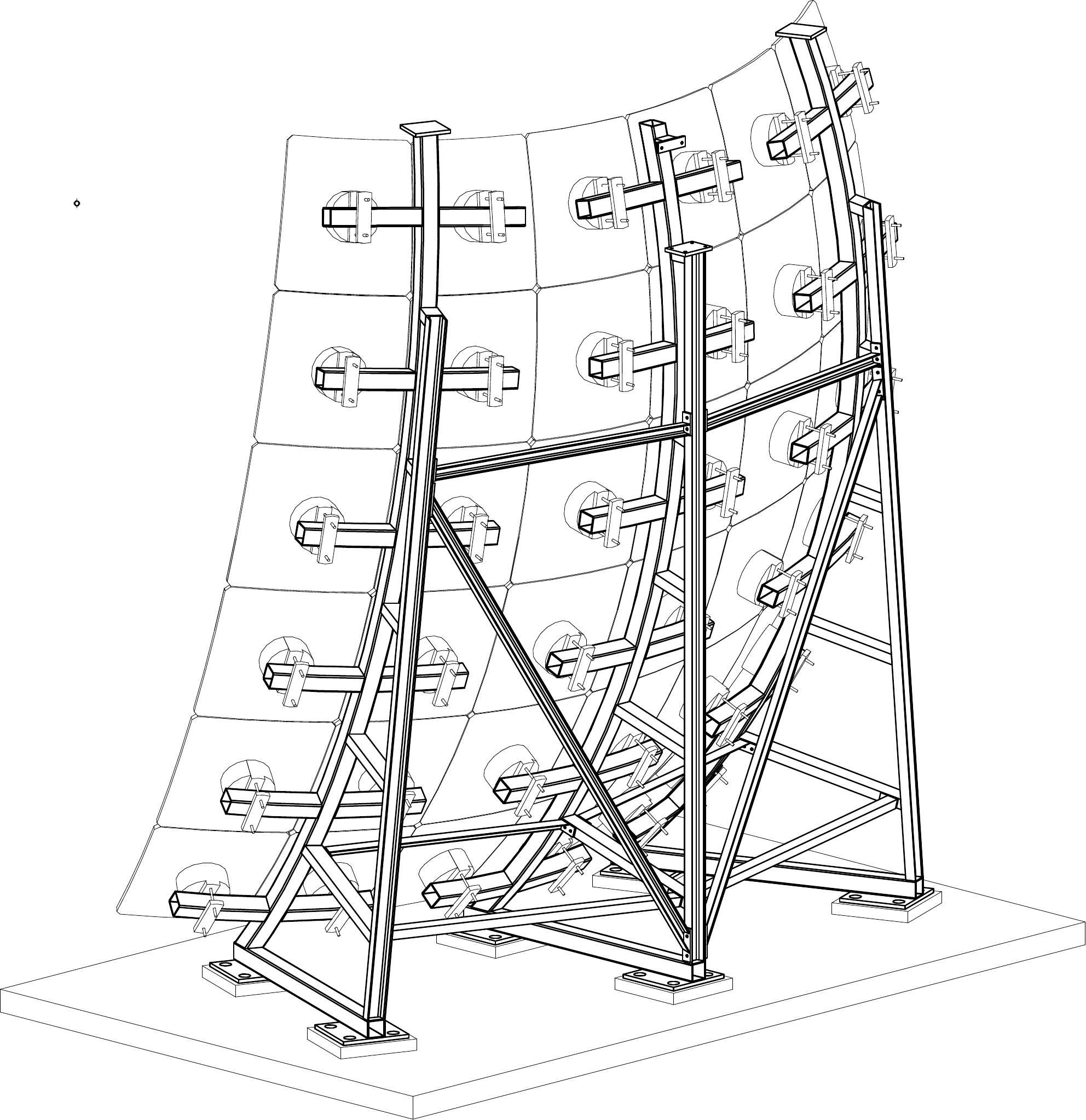}
\caption{\emph{Left:} A $6{\times}6$-segmented spherical mirror~\cite{fd}, the
same as used for the fluorescence-detector telescopes of the Pierre Auger
Observatory, where each segment is 65\,cm high with width gradually decreasing
from 65\,cm to 55\,cm. \emph{Right:} View of the support structure for the
segments at the back of the mirror. The center of the mirror sphere is also
marked with a point.}
\label{f:mirror}
\end{figure}

Here we consider HP, which are additional light U(1) gauge bosons that
kinetically mix with the (vis-\`a-vis uncharged) SM particles~\cite{Holdom:1985ag}. The
effective Lagrangian can be written as
\eq{
\mathcal{L} =
  - \tfrac14(F_{\mu\nu}F^{\mu\nu} + X_{\mu\nu}X^{\mu\nu})
  + J^\mu A_\mu
  + \tfrac{m^2}2 X_\mu X^\mu
  - \tfrac\chi2 F_{\mu\nu}X^{\mu\nu},
}
where $F_{\mu\nu}=\del_\mu A_\nu-\del_\nu A_\mu$ is the field strength tensor
of the photon and $X_{\mu\nu}$ that of the corresponding HP field, and $J^\mu$
the electric current. Large regions of the HP mass $m$ and mixing parameter
$\chi$ space are compatible with the observed DM signatures~\cite{arias,graham}
and are already searched for with many experimental methods: haloscopes,
helioscopes, and light-shining-through-a-wall methods~\cite{arias}.  Our
experiment, Finding U(1)s of a Novel Kind (FUNK), is dedicated to observe
possible candidates for HP with the dish-antenna method \cite{horns,jaeckel},
where the faint electromagnetic waves, emitted almost perpendicularly to
conducting surfaces, are focused by using a spherically shaped mirror. The HP
induced light emerging from any part of the mirror is thus gathered in the
center of the sphere where various detectors can be suitably placed. The DM
nature of the signal can be verified by observation of the expected small daily
and seasonal movements of the spot~\cite{doebrich}. A similar experiment is
currently being performed in Tokyo~\cite{suzuki,Suzuki:2015vka}.

\section{Mirror}

\begin{figure}[t]
\centering
\includegraphics[height=0.62\textwidth]{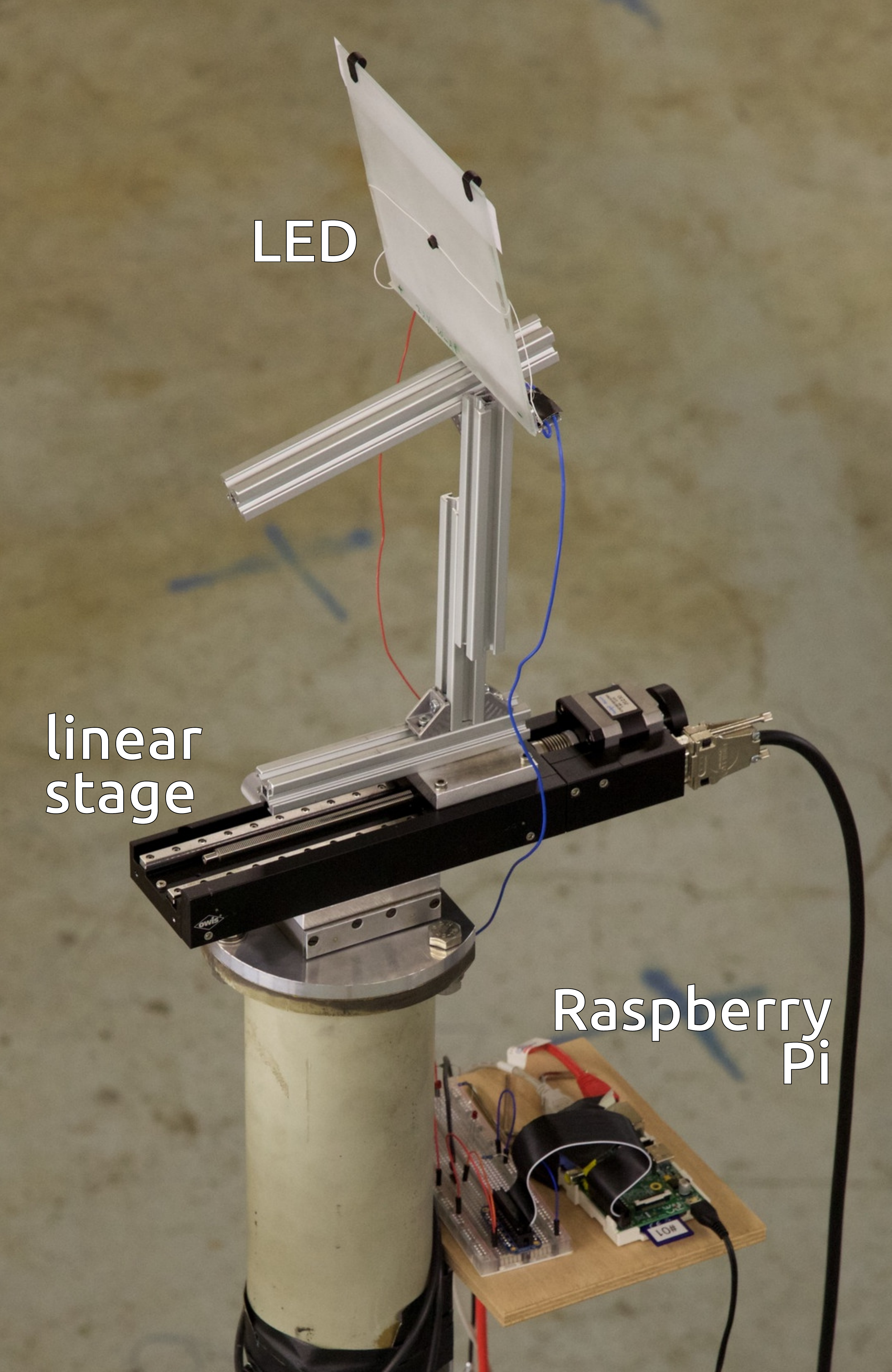}\hfil
\begin{minipage}[b]{0.475\textwidth}
\centering
\includegraphics[width=\textwidth]{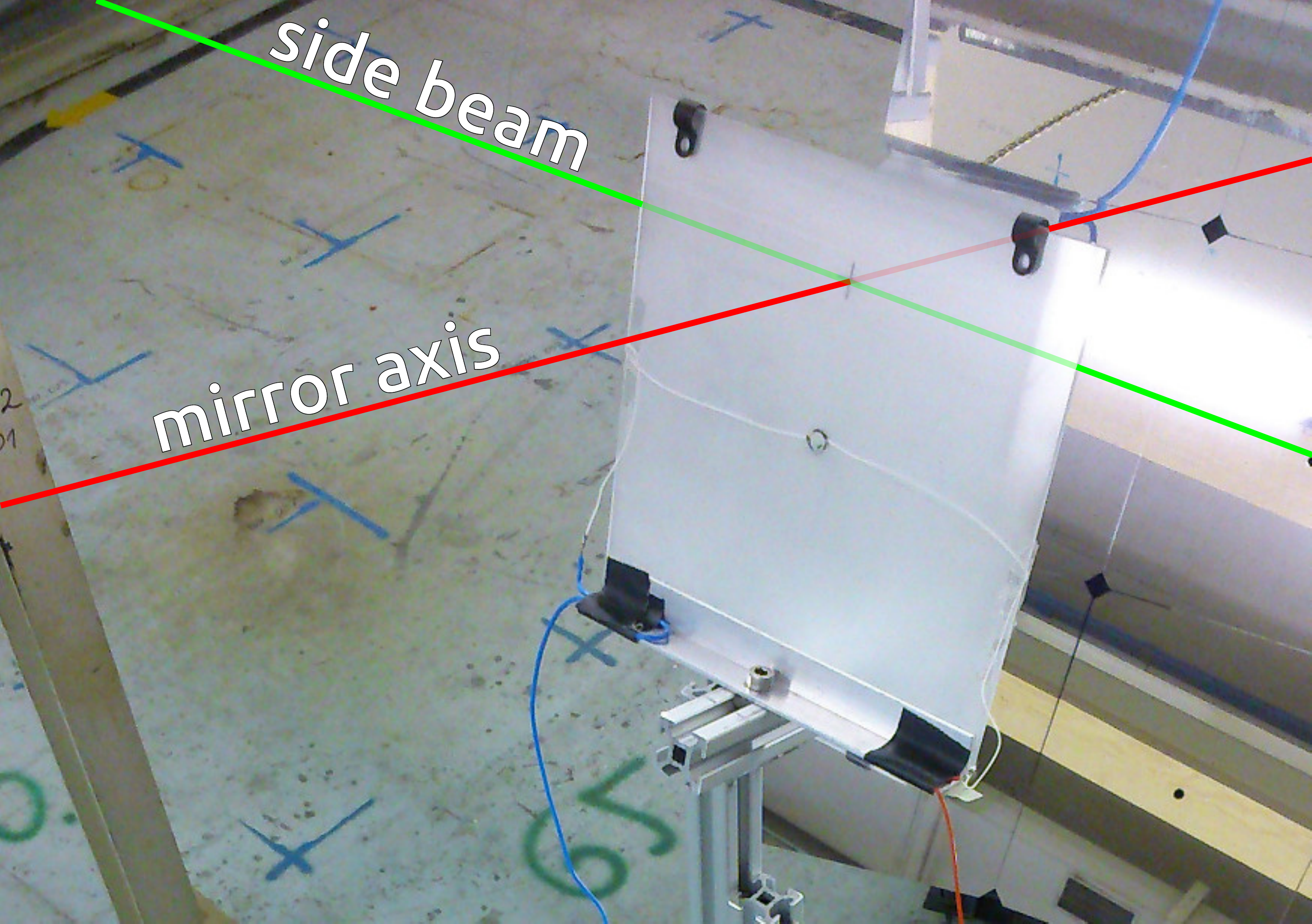}
\\[5mm]
\includegraphics[width=0.65\textwidth]{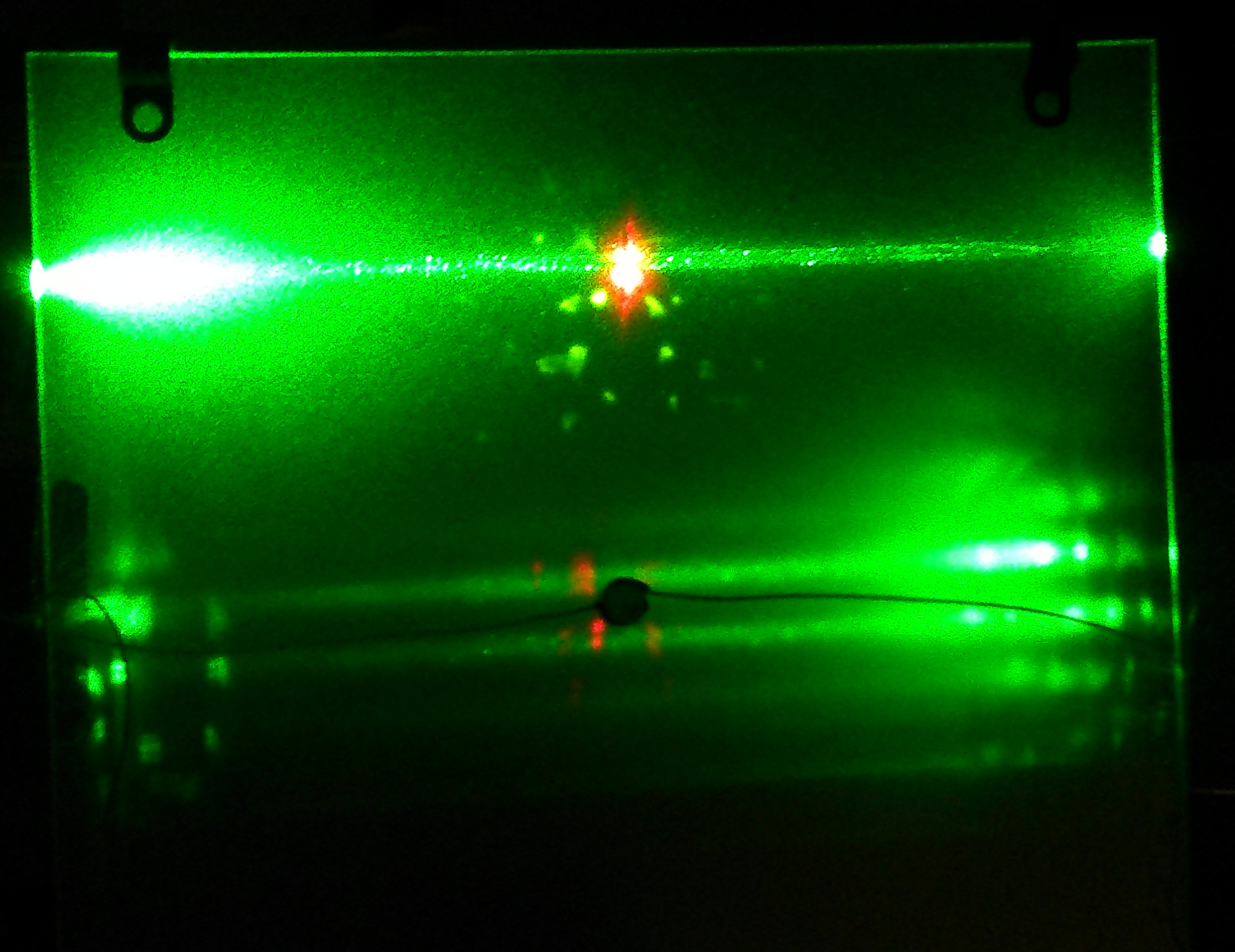}
\\[-4mm]
\null
\end{minipage}
\caption{\emph{Left:} Frosted glass screen (with a SMD LED on the front face),
linear stage with stepper motor, and a Raspberry Pi micro-computer as
controller.  \emph{Right-top:} Schematical description of the sphere-center
reference point with two crossing laser beams (red along the mirror axis, green
laterally). Mirror segments were adjusted until focused and converging into
this point.  \emph{Right-bottom:} Photo of the frosted glass and the crossing
laser beams as seen in darkness.}
\label{f:led}
\end{figure}

\begin{figure}[t]
\def\figh{0.543}
\centering
\includegraphics[height=\figh\textwidth]{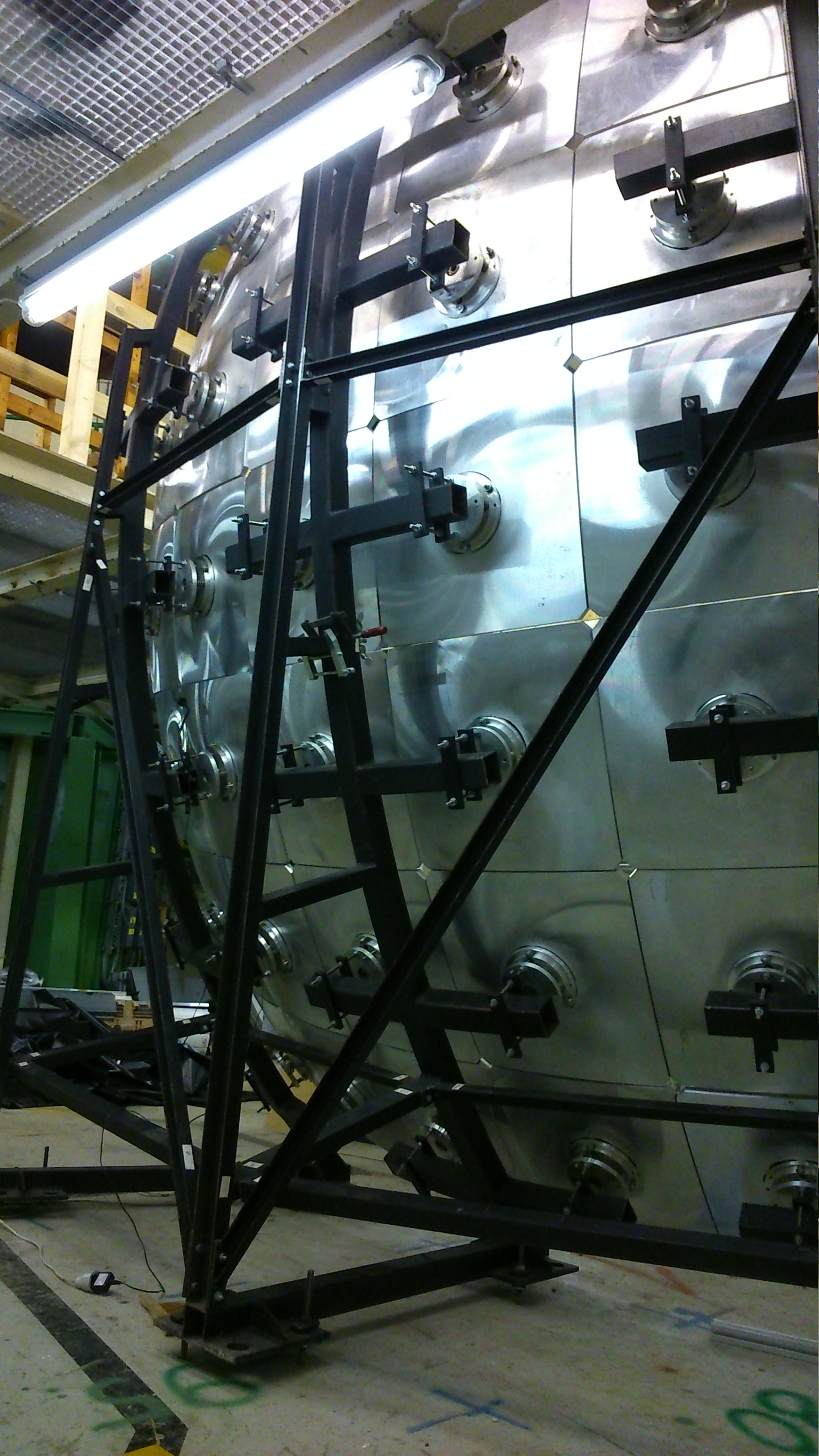}\hfil
\includegraphics[height=\figh\textwidth]{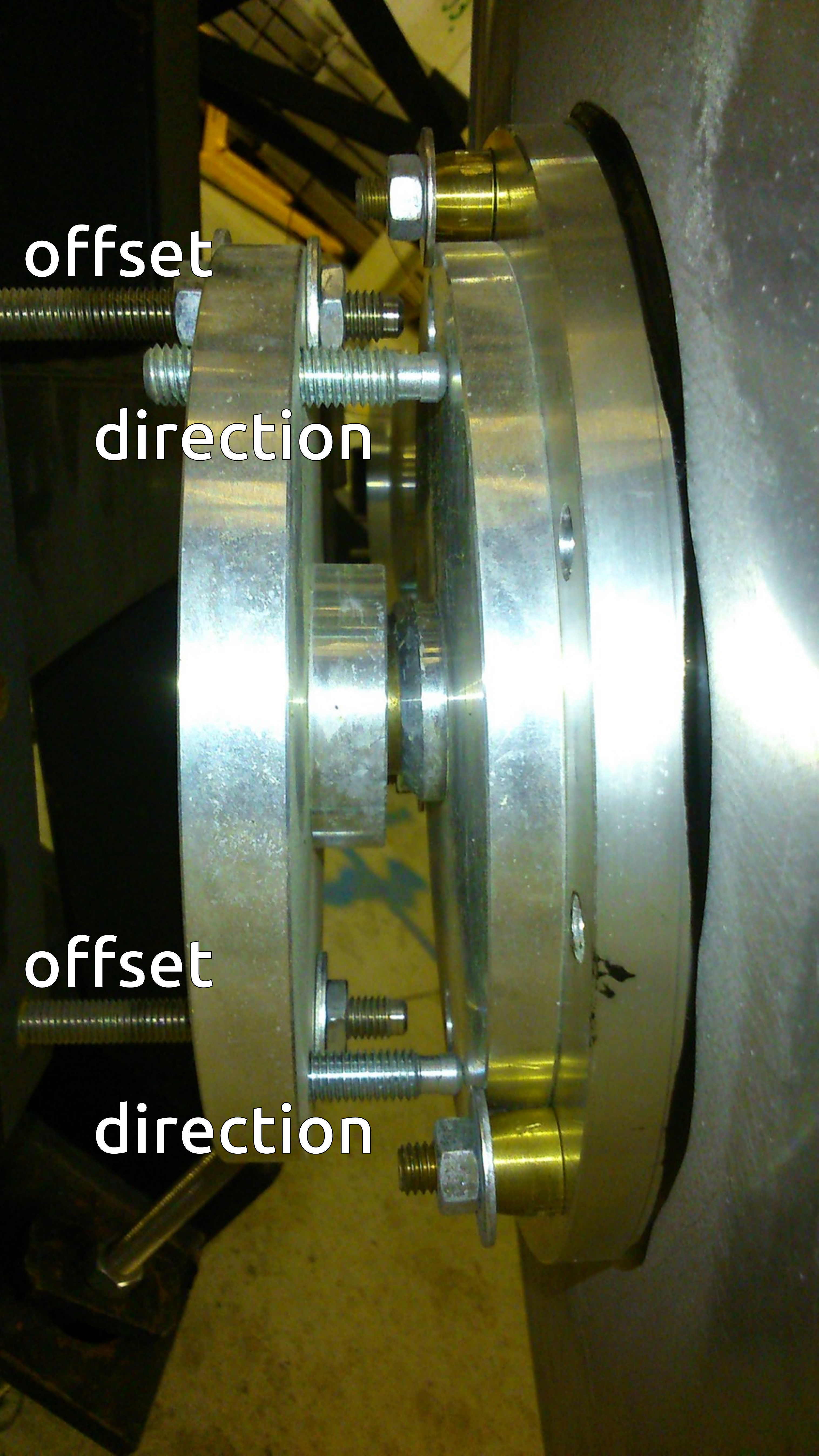}\hfil
\includegraphics[height=\figh\textwidth]{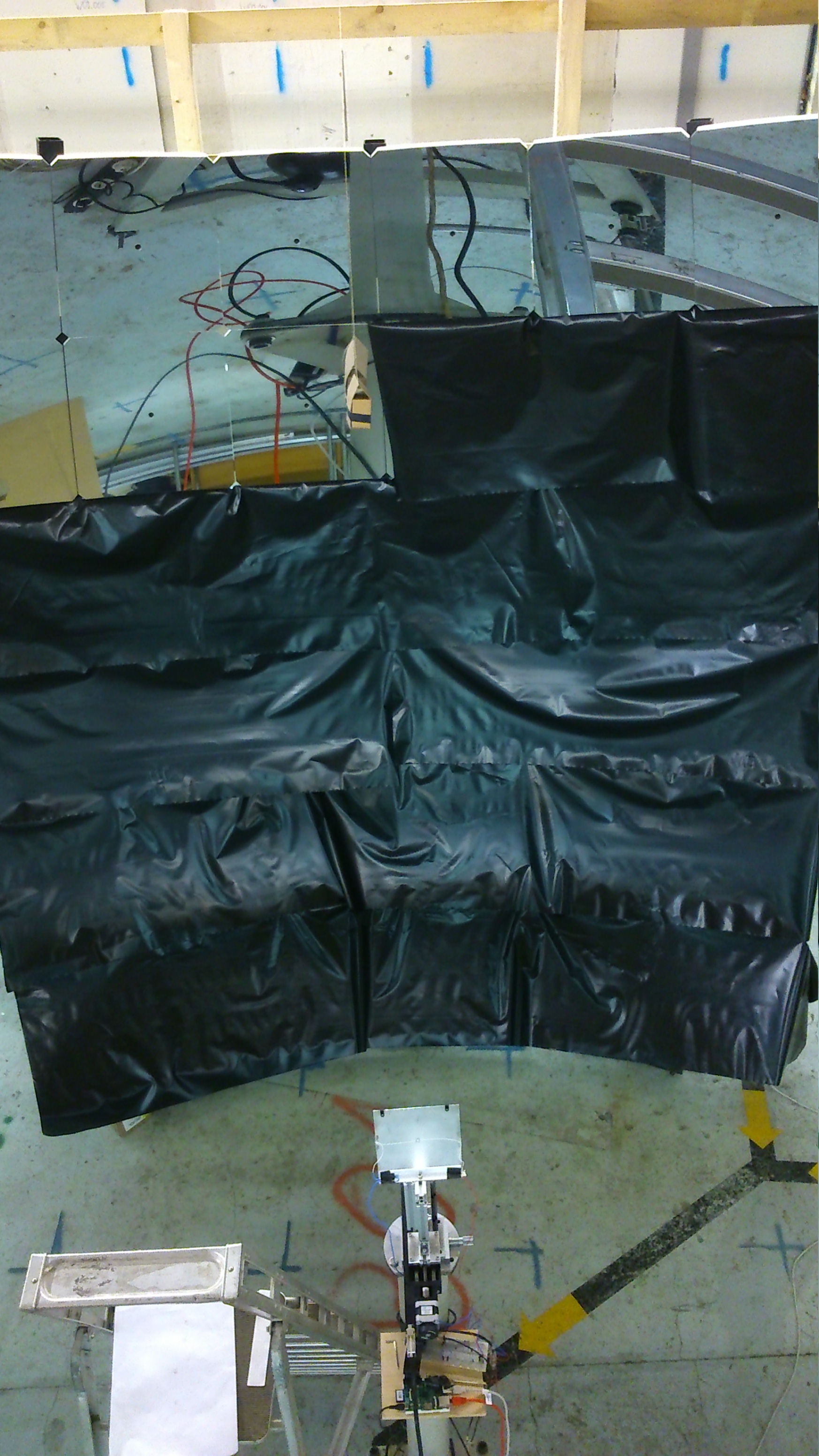}
\caption{\emph{Left:} Back side of the mirrors exposing the mounts of the
segments.  \emph{Middle:} Screws enabling longitudinal offset movement and fine
setting of the pointing direction via three-point tilting of the ball joint.
\emph{Right:} Already aligned segments are covered with non-reflecting foil to
keep only the misaligned beams in the reference point. At the bottom of the
photo the frosted-glass screen can be seen placed on a linear stage.}
\label{f:fine}
\end{figure}

\begin{figure}[t]
\centering
\includegraphics[width=0.95\textwidth]{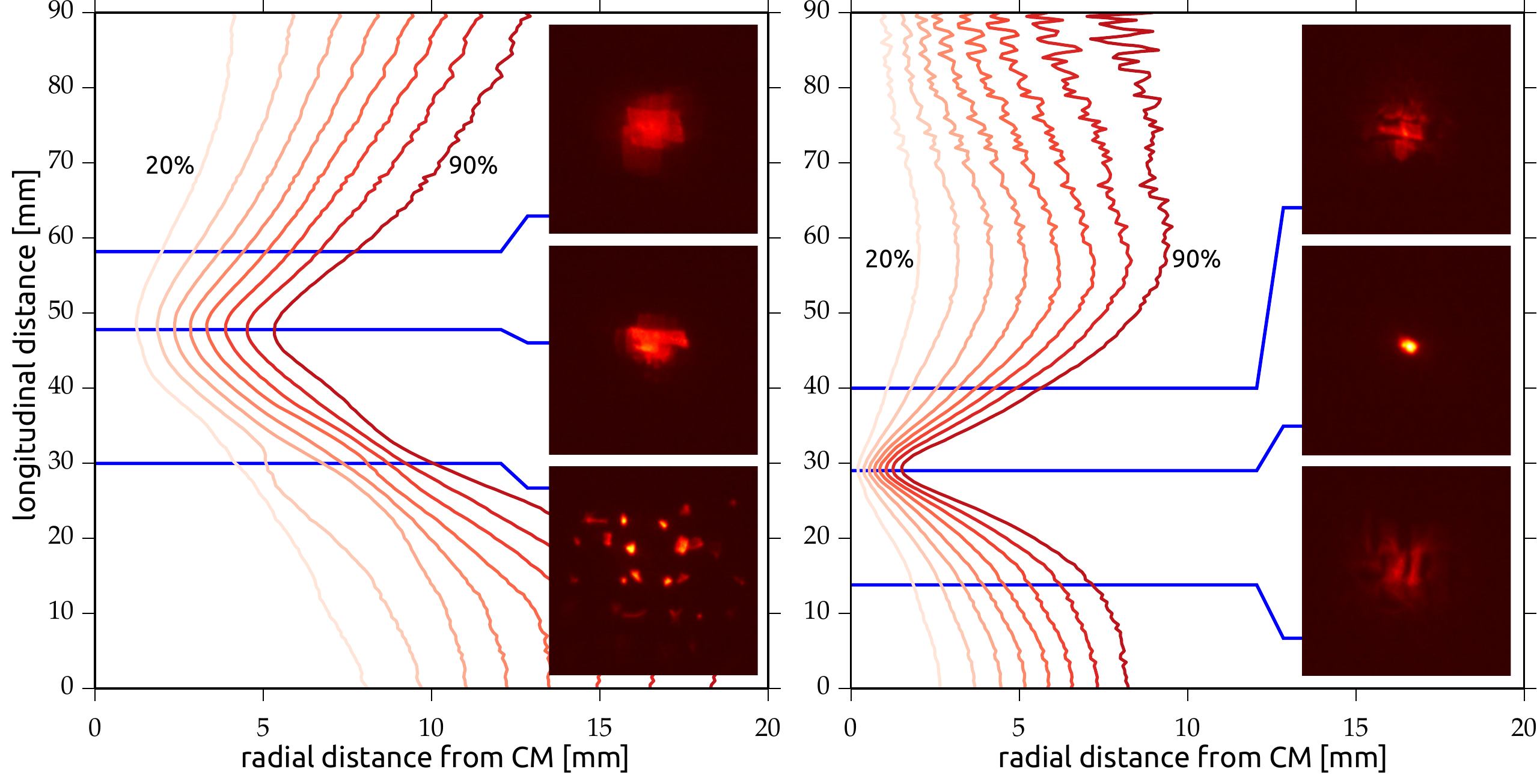}
\caption{\emph{Left:} Contours of radial distance from the image center of mass
(CM) containing certain amount of total light (in steps of 10\%) for the
initial assembly of the mirror. The three insets show spot images in various
positions marked by blue lines (middle for smallest spread). Note the focusing
of individual beams from segments but without their convergence (bottom inset).
\emph{Right:} The same as on the left but after the realignment.}
\label{f:scan}
\end{figure}

\begin{figure}[t]
\centering
\includegraphics[width=0.485\textwidth]{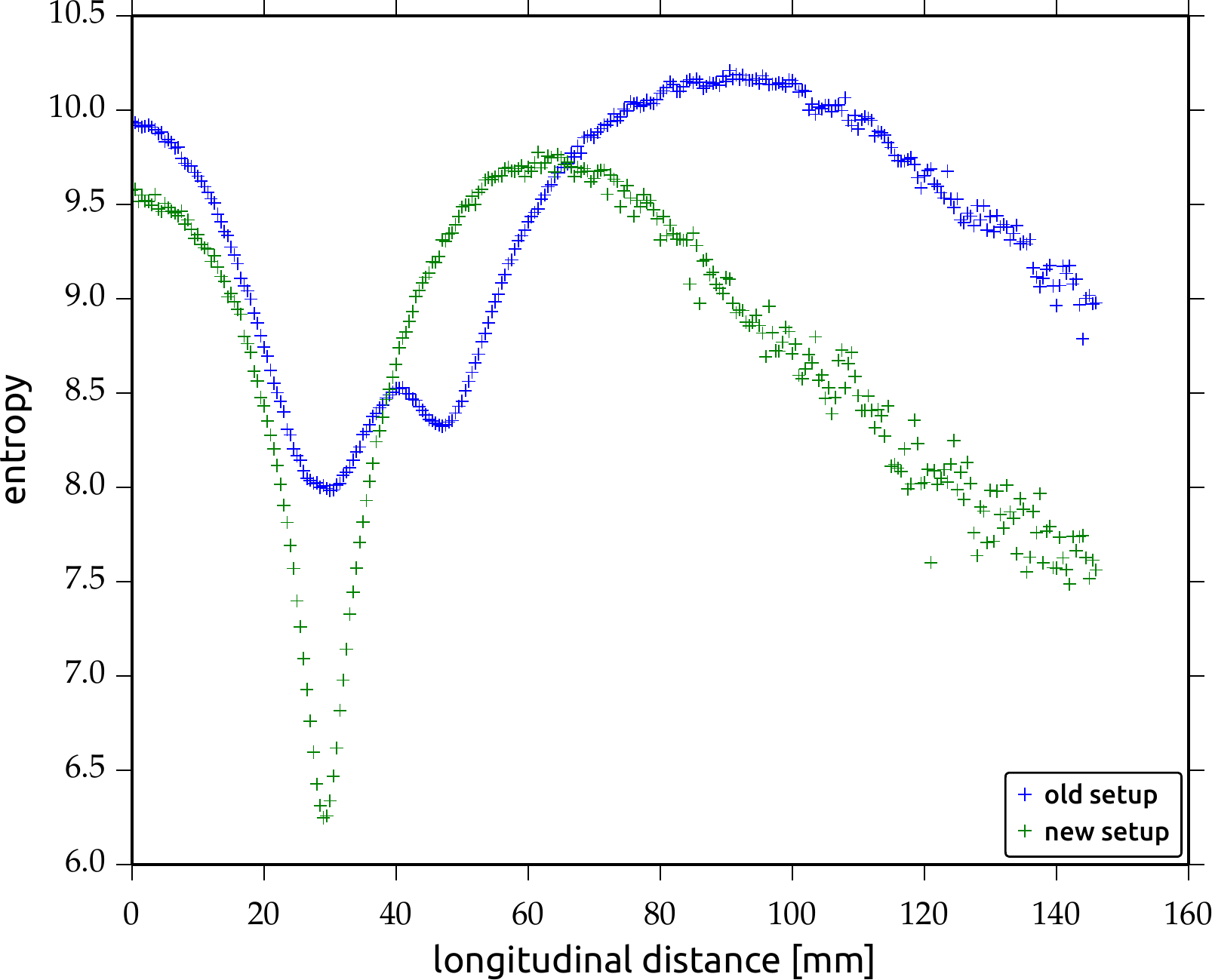}\hfil
\includegraphics[height=0.39\textwidth]{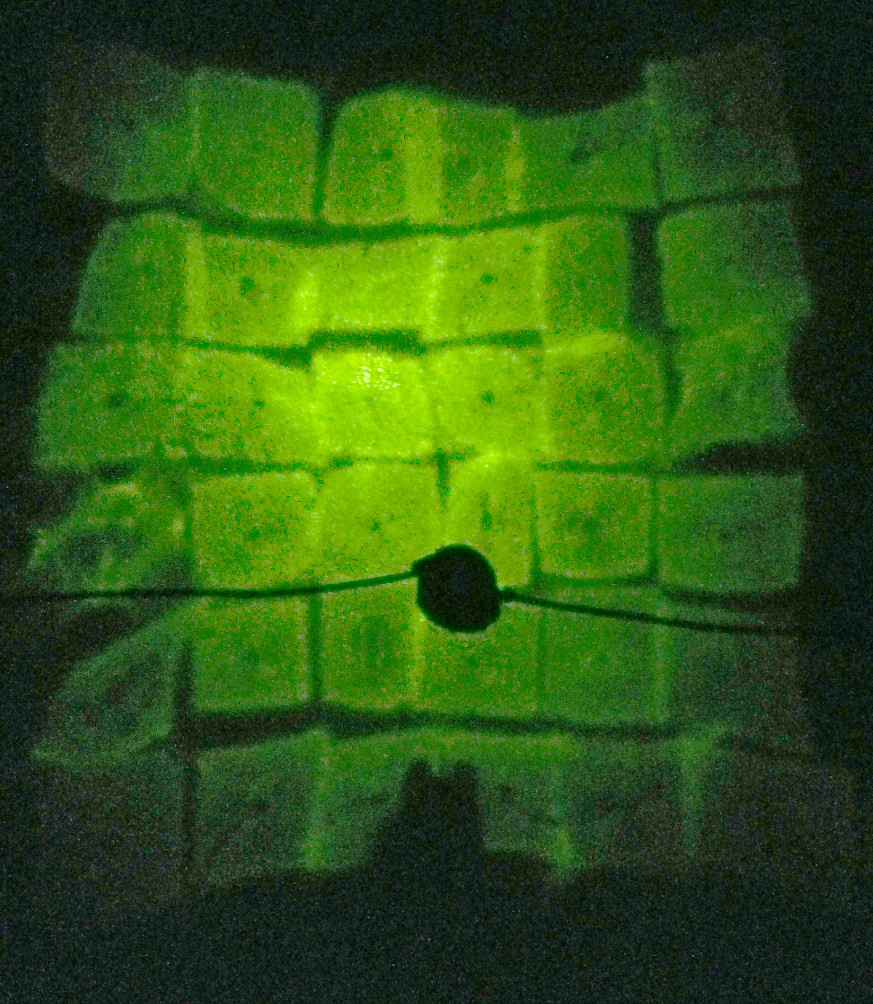}
\caption{\emph{Left:} Entropy of the image at different longitudinal positions
of the screen (blue points for the initial mirror assembly and green points
after the readjustment of individual segments). \emph{Right:} An image obtained
when the LED is driven off-center towards the mirror. Reflected beams are not
converged yet so that they form a matrix where each of the squares corresponds to
individual mirror segments, revealing their aberrations. The central black dot
is a shadow of the LED with two horizontal electricity cables.}
\label{f:entropy}
\end{figure}

For the purposes of the FUNK experiment we reused spherical mirror components
that were originally produced for the Schmidt telescopes of the fluorescence
detector of the Pierre Auger Observatory~\cite{fd}. At the KIT Campus Nord we
assembled them inside of a windowless air-conditioned experimental hall of
${\sim}20$\,m diameter with 2\,m thick concrete walls. The whole mirror is
composed of three different segment shapes that are arranged into a
$6{\times}6$ matrix (see \fref{f:mirror}). The height of the segments is
exactly 65\,cm but moving away from the mirror equator their width gradually
decreases from 65\,cm to 55\,cm.  The mirror segments were produced by milling
a cast aluminium backing to approximate spherical shape and gluing it at high
temperatures with a thinner AlMgSi sheet~\cite{ea}.  The front reflective
surface was then precisely milled with diamond tools and protected with
electrochemical anodization, resulting in a final reflectivity of
$88\pm1\%$ in visible and UV range. The segments are attached to the
mirror support structure via single-point ball joint. The extreme four corners
of the mirror form a $3.7{\times}3.1$\,m rectangle, nevertheless, due to the
spherical geometry with radius 3.4\,m, the total area of the mirror front
surface is 14.56\,m$^2$.

The initial assembly of the mirror was performed with help of an existing laser
tool according to the specifications used for the telescopes of the Auger
Observatory.  Unfortunately, the prototype components from the construction
were mirror segments which mainly did not pass strict quality requirements for
their radius of curvature or other relevant parameters. To quantify the optical
quality of the mirror we built a movable imaging platform at the center of the
mirror sphere (see \fref{f:led}-left). It consists of a frosted-glass screen
with a yellow-green SMD LED as a light source. Note that in the case of an
ideal spherical mirror, light from a point source placed exactly in the center
of the sphere will be reflected and focused back onto the source itself. With a
slight lateral offset of the source the resulting image can be observed in the
screen next to it, with the offset-related distortions being very small.  This
reflected light pattern was then observed with a CCD camera. Placing the screen
on a movable linear stage enabled us to obtain cross-section images of the
converging light beams from the mirror segments in various longitudinal
positions along the optical axis of the mirror. In \fref{f:scan} such a scan
through the central region is presented with contour lines for distances (from
the image center of mass) containing certain fractions of the total light (in
10\% steps).  As seen in \fref{f:scan}-left, the initial assembly of the mirror
produced in a minimal region (middle inset) a very large focal spot size with
radius ${\sim}6$\,mm (for 90\% of total light).  We also observed that most of
the individual segments had good focus points but at distances different than
the nominal radius (see lower inset), indicating that by realigning individual
segments we can achieve better convergence and smaller final spot size. This
position was chosen as a new dedicated mirror center and permanently marked for
future reference with two crossing laser beams (see \fref{f:fine}-right) which
will later-on also help to accurately position the measuring equipment.

In a dedicated readjustment campaign we aligned all mirror segments, either by
offsetting them longitudinally or changing their pointing direction (see
\fref{f:fine}) until the sharpest possible spot has been produced by the
corresponding segment. In this way we compensated for the non-nominal values of
the curvature while other aberrations, nicely seen in an off-center image in
\fref{f:entropy}-right, remained. Nevertheless, out of 36 segments, we
identified only 5 that had the individual spot sizes considerably larger than
the LED source size, mainly due to having two different radii of curvature in
the two perpendicular lateral directions, in \fref{f:entropy}-right identified
as having more rectangular instead of square shapes. The 90\% spot radius now
decreased to only ${\sim}2$\,mm which enabled us to use a small CCD as a
detector. Note that the seasonal oscillation of the HP spot due to the relative
movement of the mirror in the DM frame is of the same order~\cite{jaeckel}. A
way to quantify the sharpness of the image with a single observable is using
entropy, defined as $S=-\sum_i p_i\ln p_i$ where $p_i$ is the (normalized)
intensity in image pixel $i$. From the entropy an effective number of pixels or
area containing relevant signal can be defined as $N_\text{eff}=\exp(S)$. In
\fref{f:entropy}-left, entropy of the longitudinal scan images are compared.
While the initial mirror setup was exhibiting two entropy minima, a deeper one
when individual segments were focused and a shallower one for when the
individual beams crossed 20\,mm away, the entropy after the readjustment shows
only one deep minimum with at least 6 times smaller effective spot area.

\section{Preliminary results}

\begin{figure}[t]
\centering
\includegraphics[width=0.4\textwidth]{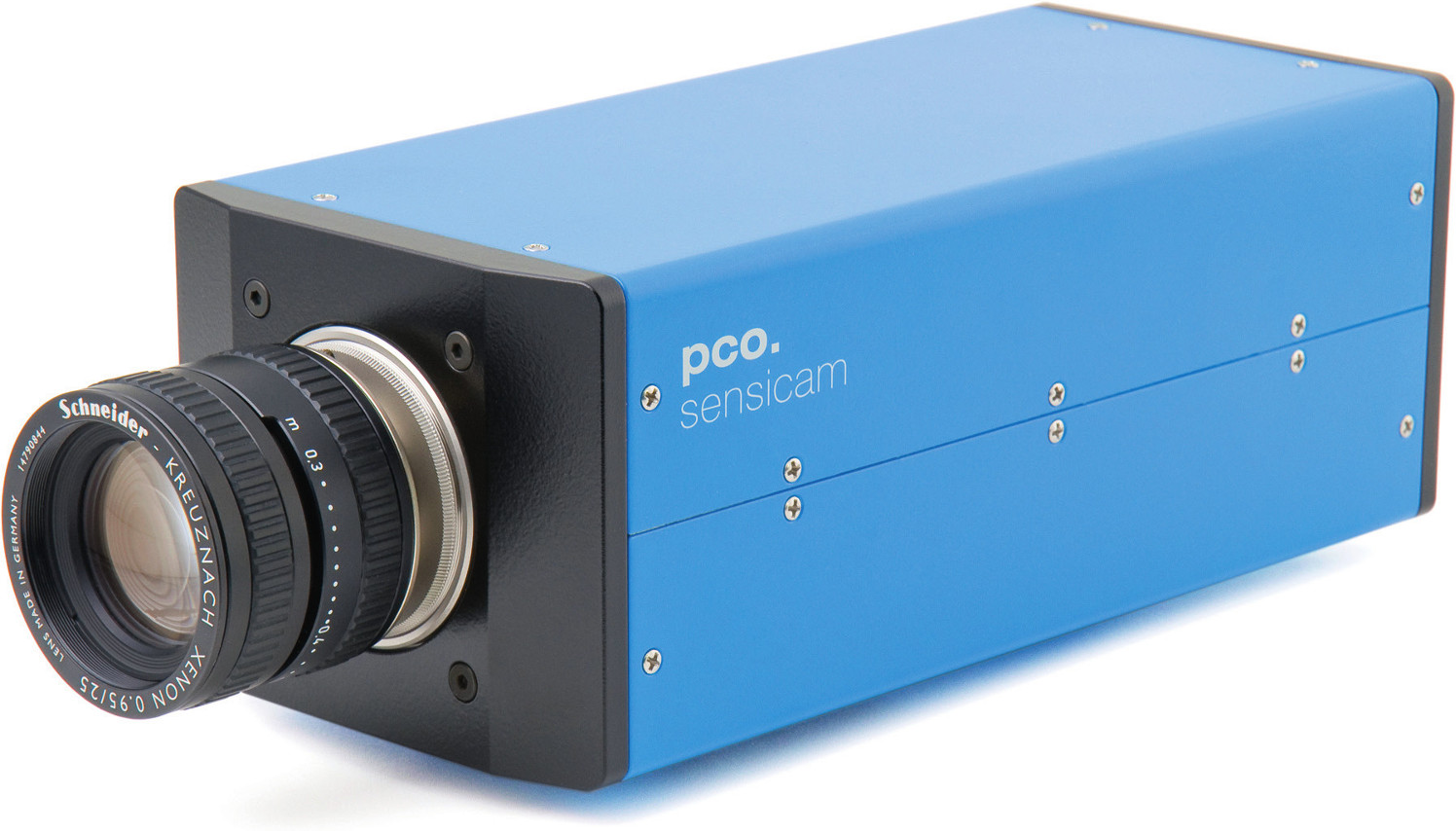}\hfil
\includegraphics[width=0.41\textwidth]{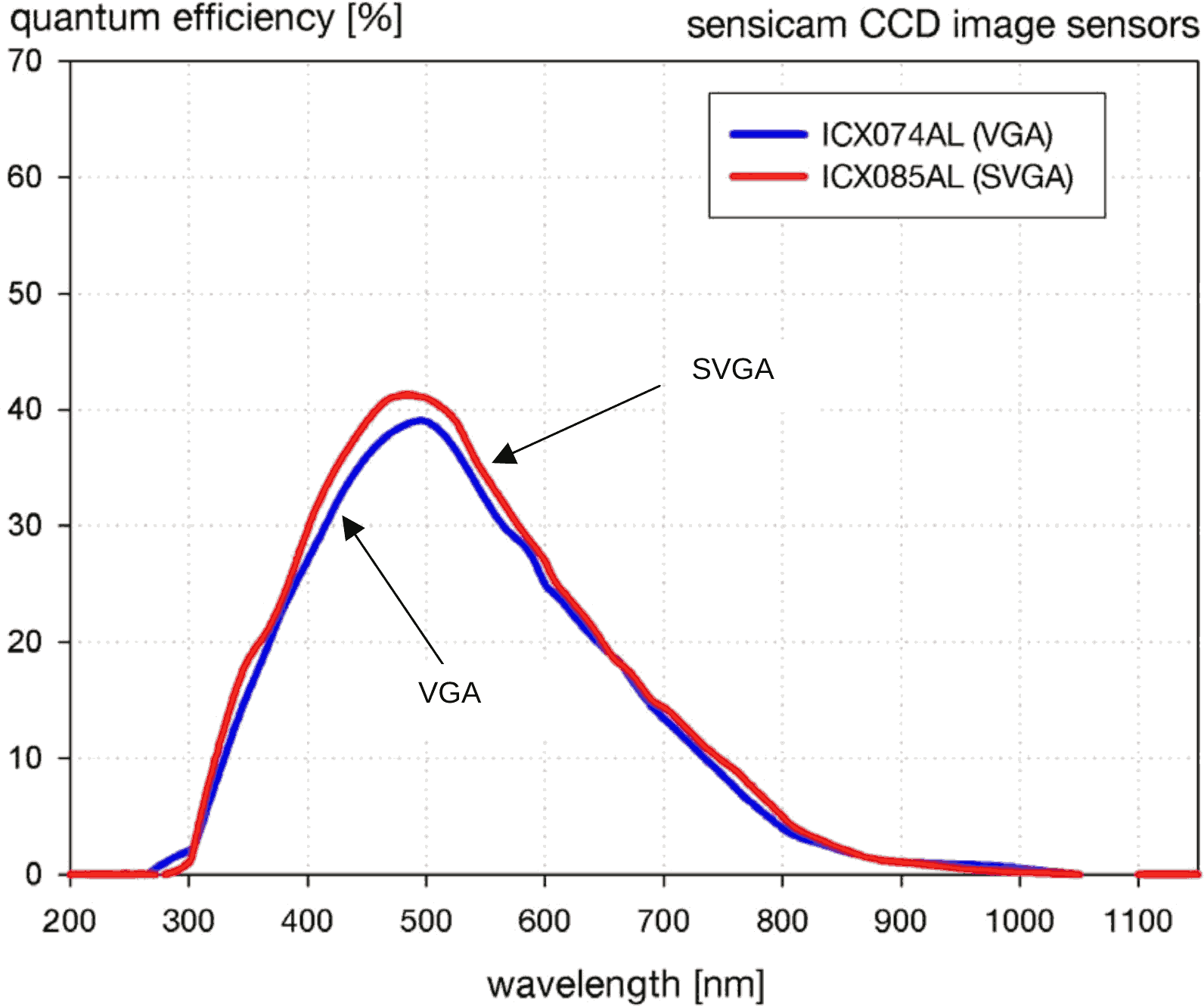}
\caption{\emph{Left:} PCO Sensicam camera used for the preliminary measurement.
\emph{Right:} Quantum efficiency of the camera CCD (see blue VGA curve) from
\cite{pco}.}
\label{f:camera}
\end{figure}

\begin{figure}[t]
\centering
\includegraphics[width=0.36\textwidth]{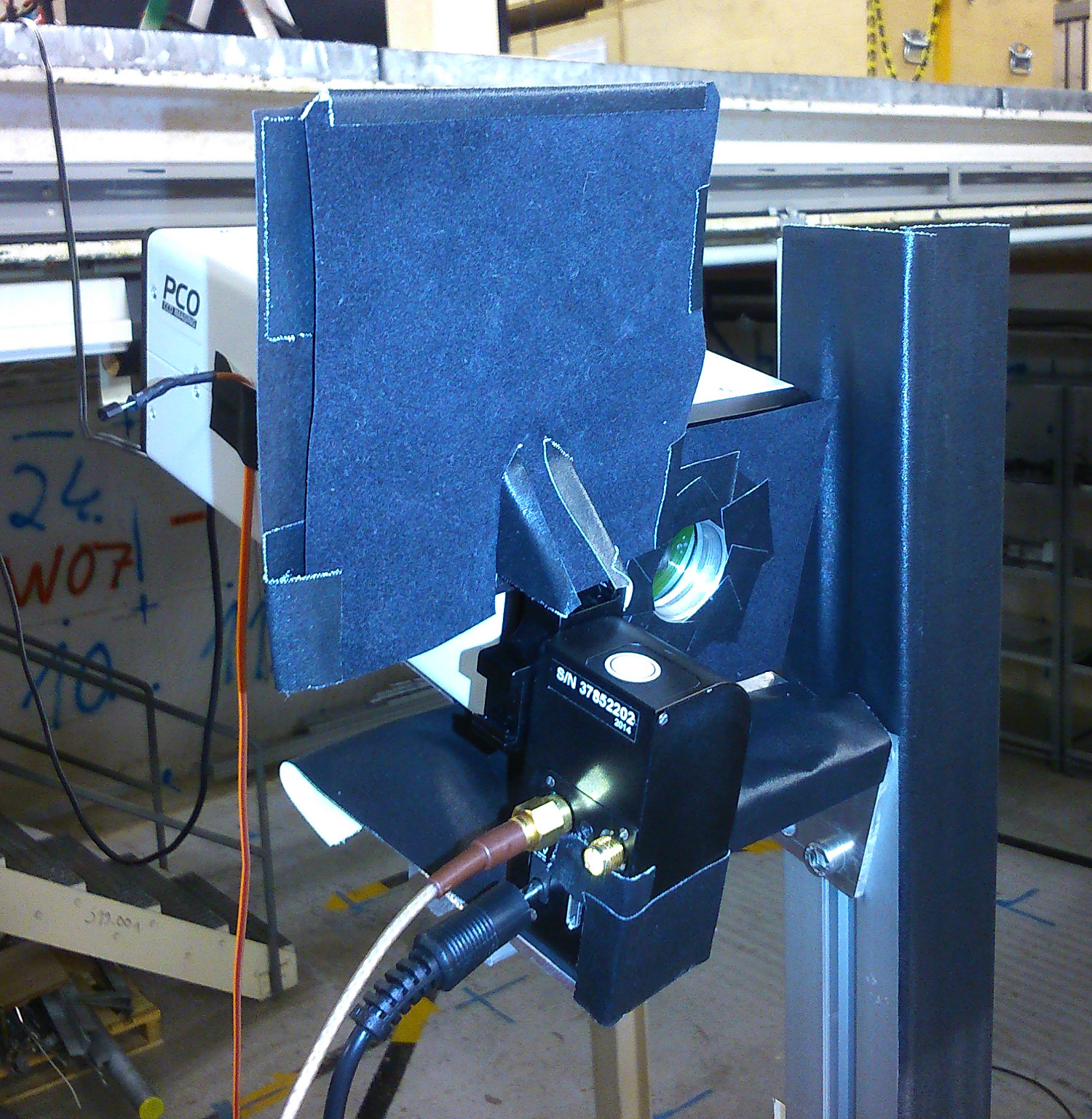}\hfil
\includegraphics[width=0.485\textwidth]{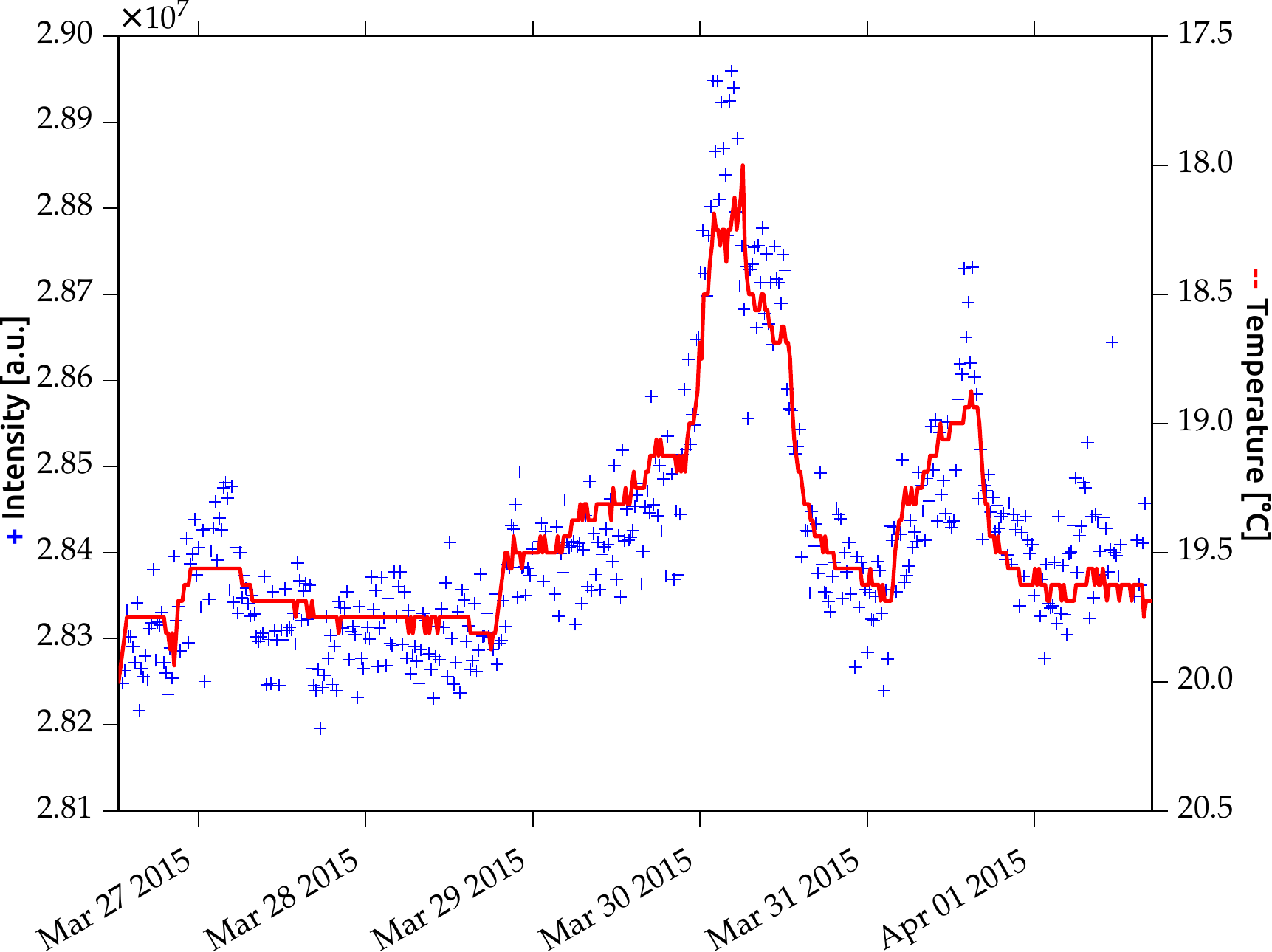}
\caption{\emph{Left:} The mounted camera with removed lens and its CCD placed
in the reference point. The Thorlabs Motorized Filter Flipper is used to drive
a shutter. \emph{Right:} The evolution of intensity (blue) with time in a
longer example run. Long term deviations are due to temperature (red)
dependence of the CCD intensity.}
\label{f:shutter}
\end{figure}

\begin{figure}[t]
\centering
\includegraphics[width=0.47\textwidth]{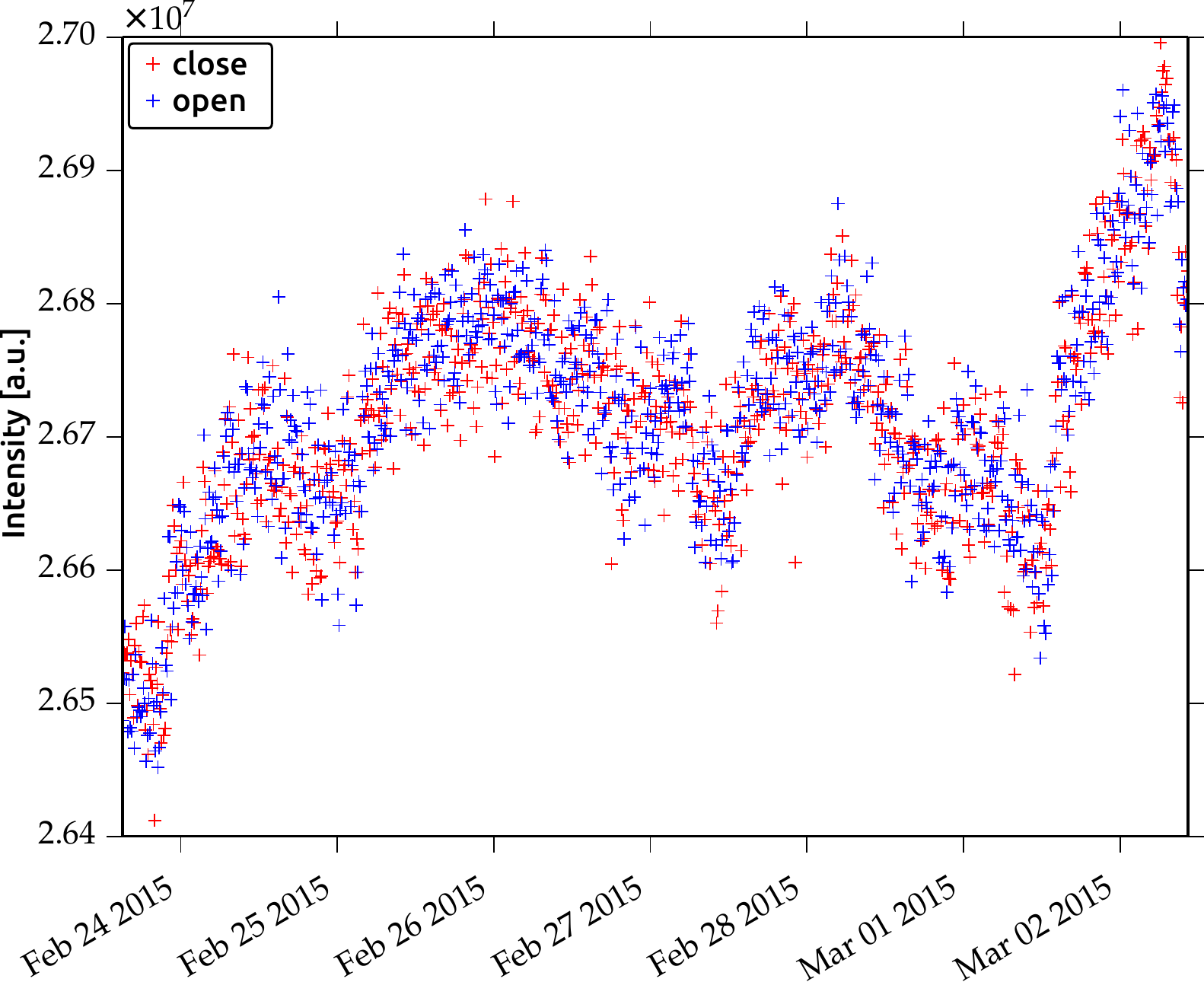}\hfil
\includegraphics[width=0.44\textwidth]{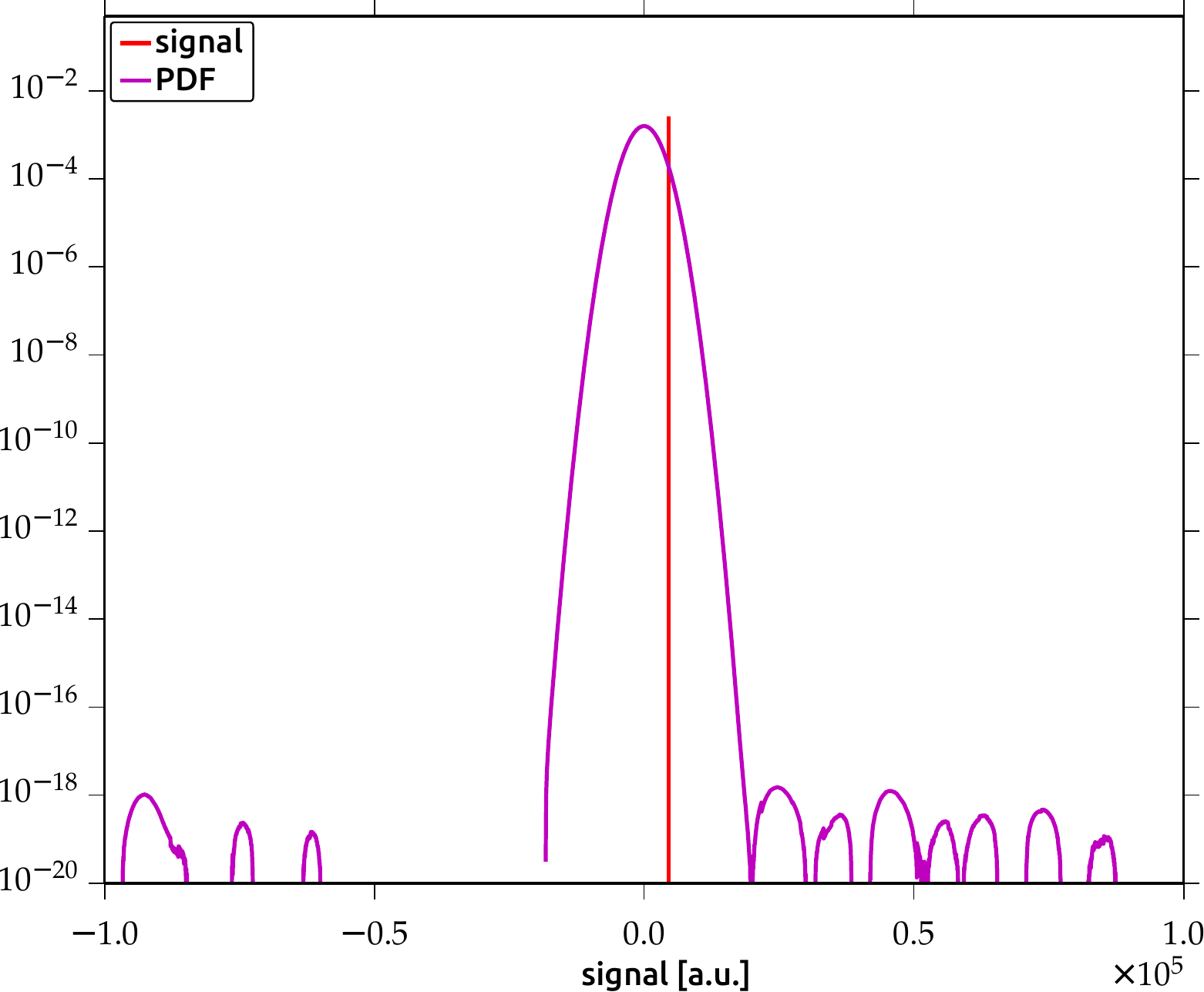}
\caption{Example of a 7 day measurement run with 1000\,s exposures.
\emph{Left:} Total image intensities for open (blue) and closed (red) shutter.
\emph{Right:} Expected distribution of intensity differences (PDF) is shown in
violet (side lobes are artifacts of the FFT convolution) and the total observed
difference of the run with red vertical line).}
\label{f:example}
\end{figure}

After the final preparation of the mirror we performed preliminary measurements
with the CCD camera that was used to obtain the images for \fref{f:scan} and
\ref{f:entropy} scans. The camera is a PCO sensicam with a half-inch diameter,
VGA resolution ($640{\times}480$) CCD sensor, internally Peltier-cooled to
$-15^\circ$C. The quantum efficiency is given in \fref{f:camera}. The collected
charge to pixel count is specified~\cite{pco} at $7.5\,e^-$/count, pixel
readout noise is ${\sim}14\,e^-$, and dark current per pixel is
${<}0.1\,e^-$/s. We removed the lens, placed the CCD into the reference point,
and added a motorized filter flipper, see \fref{f:shutter}-left, to drive a
large shutter.  Since the spot size is not much smaller than the whole CCD, the
relevant quantity for this sensor is the total intensity (integral over all
pixels of the CCD). In \fref{f:shutter}-right an example of a long-term
intensity measurement with a closed shutter is shown, together with the
temperature of the camera enclosure (red), revealing a temperature correlation
on the order of $-1$\%/K. In \fref{f:example}-left an example of image
intensities (in one of many long-term measurement runs we performed) is shown
for open and closed shutter. The signal is defined as the difference of these
two measurements. In \fref{f:example}-right a distribution of possible signals
is shown based on noise properties of individual images.  The actual signal is
shown with a red vertical bar, being perfectly compatible with the estimated
noise.

\begin{figure}[t]
\centering
\includegraphics[width=0.7\textwidth]{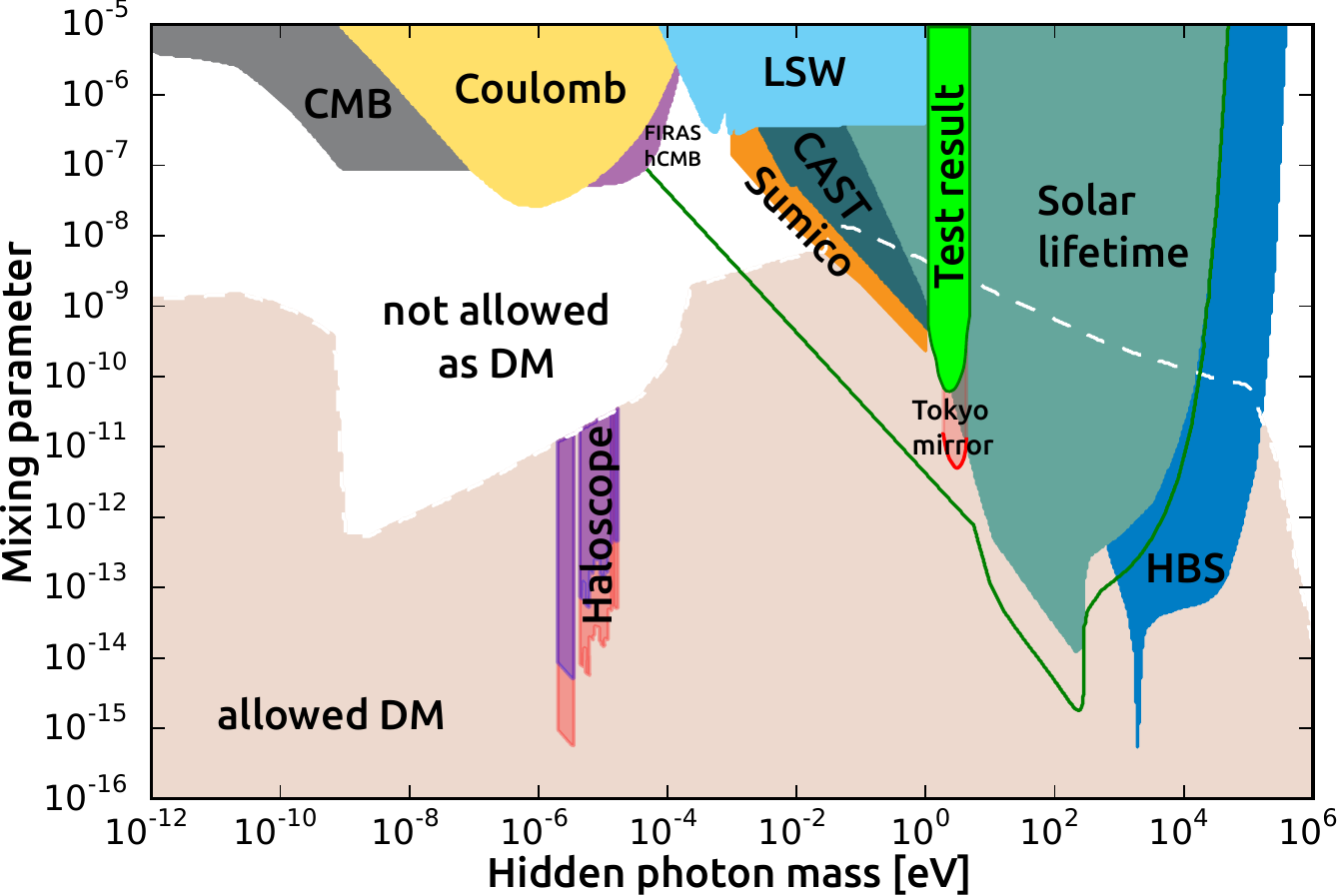}
\caption{Exclusion plot in the mixing parameter vs.\ hidden photon mass
parameter space. Meaning of some labels: measurements on cosmic microwave
background (CMB), light-shining-through-wall experiments (LSW),
horizontal-branch stars (HBS), and the green line corresponds to improved solar
limits~\cite{An:2013yfc,Redondo:2013lna}. Our preliminary result with the CCD
camera is shown in light green (Test result). Plot adapted from~\cite{suzuki}
(Tokyo mirror), for summary and labels see~\cite{goodsell}.}
\label{f:result}
\end{figure}

No signal from HP is thus observed, allowing us to set in this test example
preliminary upper limits~\cite{doebrich} on the mixing parameter $\chi$ in the
sensitivity range of the CCD, see \fref{f:result}.

\section{Conclusions and future plans}


As a first step of the FUNK experiment we assembled the segments of a prototype
mirror of the Pierre Auger Observatory to obtain a large ${\sim}14$\,m$^2$
spherical mirror, suitable for dark matter searches in the hidden photon
sector. With manual readjustments we managed to reduce the radius of the spot
size below ${\sim}2$\,mm which enables us to use small and highly sensitive
detectors of electromagnetic radiation.

While the initial tests were made with a simple CCD camera we are preparing
high-sensitivity measurements with a low dark-current CCD and a low noise PMT.
Furthermore, in the near future we want to extend the search for possible HP
mass into the MHz~\cite{aera}, GHz~\cite{crome}, and THz range.

We gratefully acknowledge partial funding from the the Helmholtz Alliance for
Astroparticle physics (HAP), funded by the Initiative and Networking Fund of
the Helmholtz Association.

\end{document}